\documentstyle[epsfig]{elsart}

\newcommand \be{\begin{equation}}
\newcommand \ba{\begin{eqnarray}}
\newcommand \ee{\end{equation}}
\newcommand \ea{\end{eqnarray}}

\begin{document}
\runauthor{Sornette and Zhou}
\begin{frontmatter}
\title{Evidence of Fueling of the 2000 New Economy Bubble by Foreign Capital Inflow:
Implications for the Future of the US Economy and its Stock Market}
\author[iggp,ess,nice]{\small{Didier Sornette}\thanksref{EM}},
\author[iggp]{\small{Wei-Xing Zhou}}
\address[iggp]{Institute of Geophysics and Planetary Physics, University
of California, Los Angeles, CA 90095}
\address[ess]{Department of Earth and Space Sciences, University of
California, Los Angeles, CA 90095}
\address[nice]{Laboratoire de Physique de la Mati\`ere Condens\'ee,
CNRS UMR 6622 and Universit\'e de Nice-Sophia Antipolis, 06108 Nice Cedex 2, France}
\thanks[EM]{Corresponding author. Department of Earth and Space
Sciences and Institute of Geophysics and Planetary Physics,
University of California, Los Angeles, CA 90095-1567, USA. Tel:
+1-310-825-2863; Fax: +1-310-206-3051. {\it E-mail address:}\/
sornette@moho.ess.ucla.edu (D. Sornette)\\
http://www.ess.ucla.edu/faculty/sornette/}
\begin{abstract}
Previous analyses of a large ensemble of stock markets have
demonstrated that a log-periodic power law (LPPL) behavior of the
prices constitutes a qualifying signature of speculative bubbles
that often land with a crash. We detect such a LPPL signature in
the foreign capital inflow during the bubble on the US markets
culminating in March 2000. We detect a weak synchronization and
lag with the NASDAQ 100 LPPL pattern. We propose to rationalize
these observations by the existence of positive feedback loops
between market-appreciation / increased-spending /
increased-deficit-of-balance-of-payment / larger-foreign-surplus /
increased-foreign-capital-inflows and so on. Our analysis suggests
that foreign capital inflow have been following rather than
causing the bubble.

We then combine a macroeconomic analysis of feedback processes
occurring between the economy and the stock market with a
technical analysis of more than two hundred years of the DJIA to
investigate possible scenarios for the future, three years after
the end of the bubble and deep into a bearish regime. We conclude
that the low interest rates and depreciating dollar are the
indispensable ingredients for a lower sustainable burden of the
global US debt structure and for allowing the slow rebuilding of
an internationally competitive economy. This will probably be
accompanied by a weak stock market on the medium term as the
growing Federal deficit is consuming a large part of the foreign
surplus dollars and the stock market is remaining a very risky and
unattractive investment. Notwithstanding strong surge of liquidity
in recent months orchestrated by the Federal Reserve, this
macroeconomic analysis which incorporates an element of collective
behavior is in line with our recent analyses of the bearish market
that started in 2000 in terms of a LPPL ``anti-bubble.'' We
project this LPPL anti-bubble to continue at least for another
year. On the short term, increased availability of liquidity (M1)
and self-fulfilling bullish anticipations may hold the stock
market for a while.
\end{abstract}
\end{frontmatter}

\section{Introduction}

According to more than a half-century of academic research, stock
market prices move almost like a (slightly biased geometric)
random walk \cite{Cootner,Malkiel}. Improvements accounting for
the presence of heavy tails in the distribution of returns and of
volatility persistence have not modified significantly the view in
the academic literature that the stock market is generally
(weakly) efficient and arbitrages away potential winning
investment strategies \cite{Famaefficient}. A more elaborated
understanding of potential gains was shown to be associated with
risk: the return-risk trade-off theory holds that there are
actually winning investment strategies but they come generally
with increasing risks, so that a larger average return is a
compensation for a larger risk \cite{Markowitz}. This constitutes
one of the central insights of modern financial economics. There
are two main versions of this return-risk trade-off. In the first
one, if a security's expected price change is positive, it may
just be the reward needed to attract investors to hold the asset
and bear the associated risks. In the second version, profits
earned by industrious investors, who are willing to work hard to
gather information and to develop winning strategies, may be
viewed as economic rents that accrue to those willing to engage in
such activities. This entails that perfectly informationally
efficient markets are an impossibility, for if markets are
perfectly efficient, the return to gathering information is nil,
in which case there would be little reason to trade and markets
would eventually collapse \cite{GrossmanStiglitz}. Black proposed
that ``noise traders'', that is individuals who trade on what they
think is information but is in fact merely noise, are the
providers of the rents \cite{Black}. More generally, at any time
there are always investors who trade for reasons other than
accruing capital gains--for example, those with unexpected
liquidity needs--and these investors are willing to ``pay up'' for
the privilege of executing their trades immediately. In
agent-based models, this competition with such ``producers'' and
the speculative investors can be shown to provide sustainable
deviations from perfect market efficiency \cite{Zhang,SlaZhang}.
Recent work in behavioral finance has also made clear that, in
order to understand market behaviors, one needs to recognize that
humans are governed by a number of non-rational considerations,
such as overconfidence, extrapolation from small samples of
evidence, loss aversion (reluctance to realize losses),
representativeness, mental accounting, relative reference levels
and self-esteem issues (reluctance to admit an erroneous
investment decision) \cite{Shefrin,Shillerexu,Shleifer}.
Professional and private investors on the stock market have not
waited for these academic developments to use so-called technical
indicators (see for instance \cite{Fosback,Bauer} and
http://www.traders.com/). A significant component of the texture
of stock market prices may thus be the result of patterns
generated by such behavioral biases and by self-fulfilling
technical indicators.

The most striking deviations from market efficiency are arguably
the speculative bubbles followed by financial crashes. The
post-mortem analysis of many financial bubbles (identified as such
after their demise) have shown that their development follow
essentially the same route. The following scenario underlying
financial bubbles and crashes has repeated itself over the
centuries and in many different locations since the famous tulip
bubble of 1636 in Amsterdam, almost without any alteration in its
main global characteristics
\cite{Galbraith,Garber,emerg,bookcrash}.
\begin{enumerate}
\item This first stage is characterized by positive economic
indicators. The bubble starts smoothly with some increasing
production and sales (or demand for some commodity), in a
relatively optimistic market. Investors form positive expectations
for the future and consequently buy the market, pushing it up.
\item The attraction to investments with good potential gains then
leads to increasing investments, possibly with leverage coming
from novel sources of funds, often from international investors
who are attracted by the potential for diversification and for
boosting their revenue. This leads to further price appreciation.
\item This in turn attracts less sophisticated investors. In
addition, leveraging is further developed with small down payment
(small margins), which lead to a demand for stock rising faster
than the rate at which real money is put in the market. In this
regime, foreign investors are even more attracted by the smaller
margin requirements and the more flexible investment rules that
are often accompanying a strong bullish market \footnote{Not all
these ingredients may be simultaneously present in a given
bubble}. \item At this stage, the behavior of the market becomes
weakly coupled or practically uncoupled from real wealth
(industrial and service) production. It is mostly the expectation
of further capital gains which continues to attract investors,
hoping to sell even higher what they buy at over-valued prices.
The bubble is in regime of self-fulfilling expectations. \item As
the price skyrockets, some investors start to cash in their gains
and they and others question the sustainability of the process. As
a consequence, the number of new investors entering the
speculative market may finally lose its momentum and the market
may enter a phase of larger nervousness, until a point when the
instability is revealed and the market collapses by the positive
feedbacks creating the selling rush.
\end{enumerate}
This scenario applies essentially to all market bubbles ending in
crashes, including old ones such as the many bubbles and bursts in
the USA during the 19th century, the bubble in the USA in the
1920s ending in the Oct. 1929 crash (the US market was considered
to be at that time an interesting ``emerging'' market with good
investment potentialities for national as well as international
investors).  In the positive sentiment accompanying and being
fueled by the developing bubble, explanations, appealing to
rational reasoning, are offered to justify the increasing prices.
The most famous one is the concept of a ``new economy,'' which are
surfaced many times, including during the bubbles in the 1920s,
during the 1960s ``tronic'' boom and during the recent bubble on
the Internet and information technology ending in the crash of
March 2000. See \cite{BAILY} for a recent survey of the
macroeconomic issues that have developed as a result of the
surprising economic performance of the 1990s expansion, including
some sense of what is known and not known about accelerated
productivity growth, the key driver of the new economy.

How can one go beyond this storytelling? A large literature has
addressed the empirical detectability of bubbles (and in
particular of ``rational expectation bubbles'') in financial data
(see \cite{Camerer,Adam} for surveys of this literature).
Empirical research has largely concentrated on testing for
exponential trends in the time series of asset prices and foreign
exchange rates \cite{Evans,Woo}. However, it has been shown that
tests for exponential roots are quite unreliable in the presence
of the more realistic variant of periodically collapsing rational
expectation bubbles \cite{Evans2}. In fact, certain variants of
bubble dynamics might look like the outcome of a stationary
process \cite{Meese}. They would, therefore, fool the testing
strategy recommended by Hamilton and Whiteman \cite{Hamilton} of
comparing the stationarity properties of both asset prices and
observable fundamentals. In a nutshell, the problem is that it is
hard if not impossible to distinguish an exponential explosive
bubble from an exponentially growing fundamental price. The
following quote from Federal Reserve Chairman A. Greenspan
illustrates a now famous view \cite{Greenspan}: `` We, at the
Federal Reserve, recognized that, despite our suspicions, it was
very difficult to definitively identify a bubble until after the
fact, that is, when its bursting confirmed its existence.
Moreover, it was far from obvious that bubbles, even if identified
early, could be preempted short of the Central Bank inducing a
substantial contraction in economic activity, the very outcome we
would be seeking to avoid.''

A different line of attack has been suggested, based on an analogy
with so-called critical phenomena in the statistical physics sense
of critical phase transitions (see
\cite{SJB96JPI,CriCrash99,JSL99,CriCrash00,SJ01QF,bookcrash} and
references therein). Two hallmarks of speculative bubbles have
been documented: (i) super-exponential power law acceleration of
the price towards a ``critical'' time $t_c$ corresponding to the
end of the speculative bubble and (ii) log-periodic modulations
accelerating according to a geometric series signaling a discrete
hierarchy of time scales. Out-of-sample tests \cite{endoexolppl}
have shown that, over many analyzed world markets in which 49
drawdown outliers were identified, the log-periodic power law
(LPPL) precursors are present in 25 cases, qualifying these
bubbles as endogenously generated. Restricting to the world market
indices, 31 drawdown outliers have been identified of which 19 are
preceded with clear LPPL precursors. Systematic non-parametric
tests have also confirmed the relevance of LPPL structures for the
detection of crashes or severe changes of regimes
\cite{nonparacrash,szpat}.

All these studies have been focused on financial time series taken
one at a time, to test whether past market anomalies could be
associated with incoming market instability. But, of course, the
world is multi-dimensional and especially the financial world is
multi-variate: interest rates, implied volatility, volumes,
exchange rates, cross-sectional dependencies between equity and
commodity time series should in principle all be integrated in a
global analysis of speculative bubbles. One step in this direction
was performed for the analysis the bearish market trends
unraveling since the summer of 2000 \cite{SZ02QF,ZS02}, which
showed that the majority of European and Western stock market
indices as well as other stock indices exhibit practically the
same LPPL ``anti-bubble'' \footnote{An ``anti-bubble'' is defined
as a self-reinforcing generally severely declining price
trajectory with self-similar expanding log-periodic oscillations,
which are the symmetric to LPPL bubbles under a time reversal.}
structure as found for the USA S\&P500 index.

Here, we present new evidence in favor of the LPPL theory of
speculative bubbles by testing quantitatively one of the
predictions of the scenario $1-5$ presented above, which underlies
financial bubbles and crashes. This prediction concerns the role
of foreign investors in providing increasing amounts of funds,
therefore driving up the stock market of the target country as the
bubble develops. Many researchers have indeed stressed the impact
of foreign investments in the fueling of bubbles, preparing the
stage of financial instabilities and crashes \cite{Galbraith}.
Albuquerque et al. have analyzed the unparalleled increase in
foreign direct investment to emerging market economies of the last
25 years \cite{Albu}. Kim and Wei have shown that foreign
investors outside Korea before and during the Korean currency
crisis were more likely to engage in positive feedback trading
strategies and were more likely to engage in herding than the
branches/subsidiaries of foreign institutions in Korea or foreign
individuals living in Korea \cite{KimWei}. Foreign investors'
strategies on future and option contracts have been found to have
a destabilizing role \cite{Ghysel}. Dahlquist and Robertsson found
a strong link between foreigners' trading and local market returns
\cite{Dahlquist}. They also found that foreigners act like
non-informed feedback traders as they increase their net holding
in firms that have recently performed well. On the other hand,
analyzing the effects of the 1987 stock market crash on the London
markets, the bursting of the Japanese stock market and real estate
market bubble in 1989-1990, the short term and long term
consequences of the turbulence in the European exchange rate
system in 1992 and 1993, the short term and long term effects of
the Mexican devaluation in 1995 for Mexico and other emerging
stock markets and the developments in South-East Asia in
1994-1997, Tesar and Werner purport that, while foreign investors
at times set a financial market crisis in motion (especially
currency crises), they have not been the underlying cause of the
crises nor have they exacerbated these crises \cite{Tesar}. Dooley
and Walsh survey the large literature on capital flows and discuss
several possible origins of induced destabilization \cite{Dooley}.

\begin{figure}
\begin{center}
\epsfig{file=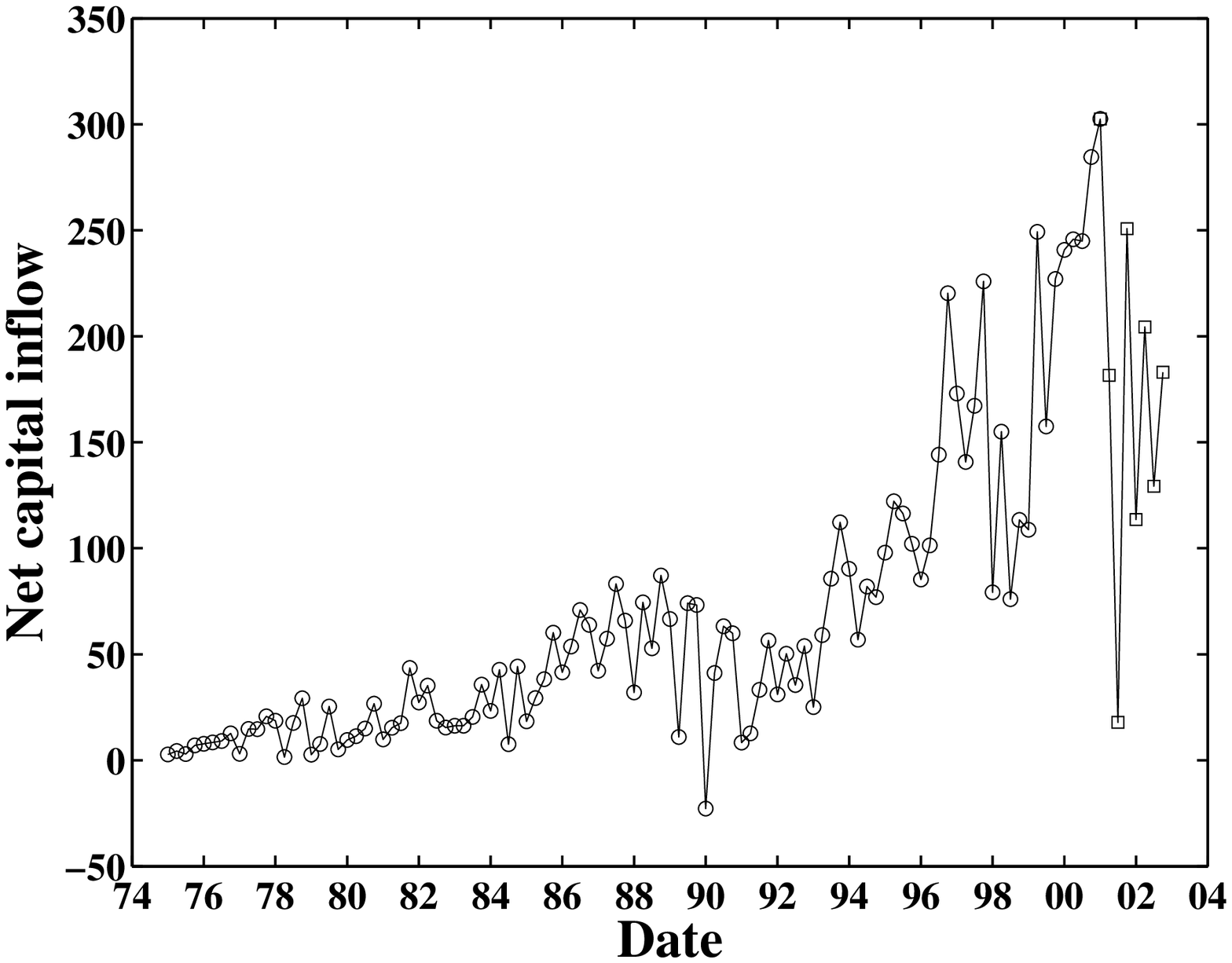,width=12cm, height=10cm}
\end{center}
\caption{Time evolution of the foreign net capital
inflow $I(t)$ to the USA per quarter since 1975. The unit of
$I(t)$ is in billion of US dollars while time $t$ is in calendar year.} \label{Fig:It}
\end{figure}

We propose to add on this literature by analyzing the relationship
between the stock market in the late 1990s and the net foreign
capital inflows in the USA, available from the Federal Reserve's
web site\footnote{FRED II: http://research.stlouisfed.org/fred2/}.
The source is the U.S. Department of Commerce, Bureau of Economic
Analysis. The data correspond to the net flux of US international
transactions: investments of foreigners in the US are counted
positive and US investments outside are deducted. The net flux
corresponds to a given quarter and is seasonally adjusted. The
time evolution of the US foreign assets net capital inflow $I(t)$
per quarter since 1975 is shown in Fig.~\ref{Fig:It}. The net
inflow increased over the years, accelerating to very large values
before crashing in the first quarter of 2001.

\section{Testing for a LPPL signature in the foreign capital inflow}
\label{s1:pre}

\subsection{Log-periodic power law accelerating structure}
\label{s2:PLLP}

We follow the methodology explained in our previous papers
\cite{SJ01QF,bookcrash,endoexolppl,nonparacrash}. Specifically, we use
the generalization of the LPPL formula in terms of
the truncated second-order Weierstrass-type function \cite{GS02PRE}
\begin{equation}
I(t) = A + B x^{m} + {\Re{\left(\sum_{n=1}^2 C_n
{\rm{e}}^{i\psi_n}x^{-s_n}\right)}}~, \label{Eq:It}
\end{equation}
where $x= t_c-t$ with $t_c$ the end of the bubble and
${\Re{(\cdot)}}$ stands for the real part. The phases $\psi_1$ and
$\psi_2$ are determined from the fit to the data. The exponent
$s_n$ is defined as $s_n = m + i n \omega$. Its real part is the
exponent $m$ which controls the overall power law dependence of
the price trajectory. Its imaginary part $\omega$ is the angular
log-frequency of the log-periodicity. The component $n=2$ allows
for the existence of an harmonics.

This second-order Weierstrass-type function (\ref{Eq:It}) proved
to outperform our former specification in terms of a first-order
LPPL function for modelling of the worldwide 2000-2002
anti-bubbles that started in mid-2000 \cite{SZ02QF,ZS02}. The
special case $C_n = C/n^{m+0.5}$ recovers the zero-phase
Weierstrass-type function, which captures well the five successive
crashes within the 2000-2002 anti-bubble in the US S\&P 500 index
\cite{ZS03} and models well the ongoing UK real estate bubble
\cite{house}.

We fit the seasonally adjusted quarterly net capital foreign
inflow to the USA $I(t)$ of foreign assets (transactions over a
given quarter) from 1975 to the first quarter of 2001. We use the
algorithm and the fitting procedure described in \cite{ZS03}. The
optimization search was performed for a critical time $t_c$ in a
wide range of $t_{\rm{last}}\pm 1000$ trading days by replacing
$t_c-t$ by $|t_c-t|$ in fitting \cite{SZ02QF}, where
$t_{\rm{last}}$ is the date of the last point used in the fitting
analysis.

\begin{figure}
\begin{center}
\epsfig{file=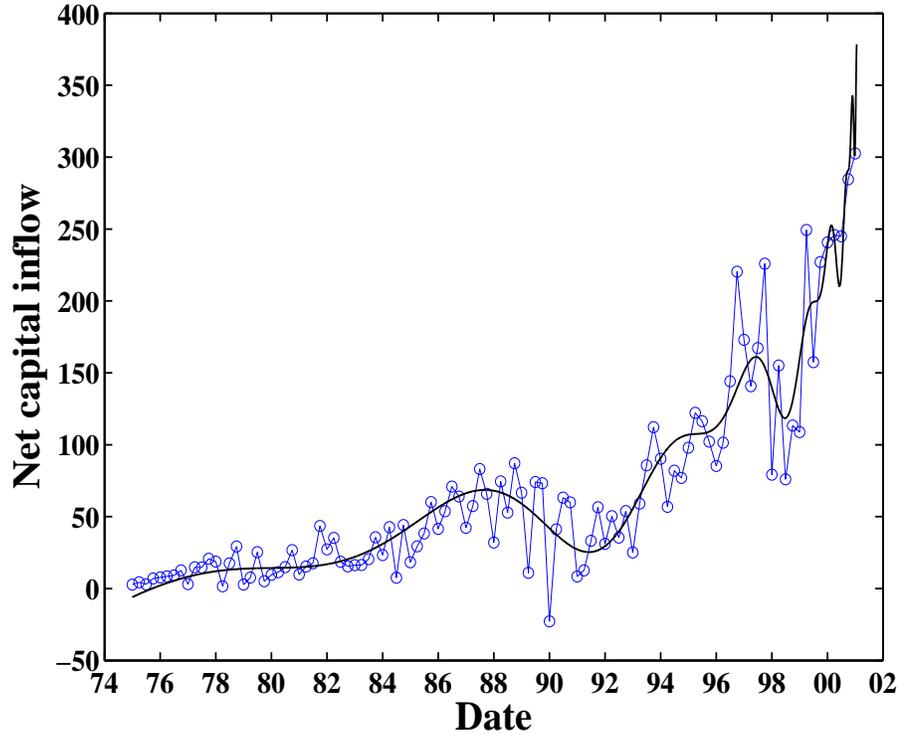,width=12cm, height=10cm}
\end{center}
\caption{Fit of the time evolution of the foreign net capital
inflow $I(t)$ in the USA from 1975 till
the first quarter of 2001 when it reached its maximum, by
a second-order
Weierstrass-type function given by expression (\ref{Eq:It}). The predicted critical
time is $t_c={\rm{2001/03/12}}$, the power-law exponent is
$m=0.01$, and the angular log-frequency is $\omega=4.9$. The
fitted linear parameters are $A=7355$, $B=-6719$, $C_1=21.5$ and
$C_2=16.2$. The r.m.s. of the residuals of the fit is 22.810.}
\label{Fig:BOPI}
\end{figure}

Figure \ref{Fig:BOPI} shows the fit of the net inflow $I(t)$ to
Eq.~(\ref{Eq:It}). The predicted critical time is
$t_c={\rm{2001/03/12}}$, which is consistent with the real date of
the inflow crash that occurred in the first quarter of 2001. Four
well-developed log-periodic oscillations can be observed in
Fig.~\ref{Fig:BOPI}. We can also identify their second-order
harmonics, which we analyze further below. We have constrained the
formula to have a non-negative exponent $m$, to ensure that $I(t)$
remains finite at $t_c$. With this constraint, the fit chooses the
lowest $m=0.01$ that we allow. Slightly better but not
significantly better fits are obtained by freeing $m$, which then
adjusts to values around $-0.5$, while $t_c$ is moved to
2002-2003. A negative exponent would lead to an unrealistic
divergence of the flow and we thus reject this family of solution.
This appears to be justified by the also unrealistic values of
$t_c$ obtained with negative $m$ values. Such a small value of the
exponent $m$ means that the power law is equivalent to a
logarithmic law $\sim \ln (t_c-t)$, as proposed in
\cite{Van1,Van2,Van3}. The other parameters of the fit are
$\omega=4.9$, $A=7355$, $B=-6719$, $C_1=21.5$ and $C_2=16.2$. The
r.m.s. of the fit residuals is $22.810$. Note that the comparable
amplitudes of $C_1$ and $C_2$ imply the existence of a significant
second-order harmonic in the log-periodic structures of the net
inflow bubble.

\begin{figure}
\begin{center}
\epsfig{file=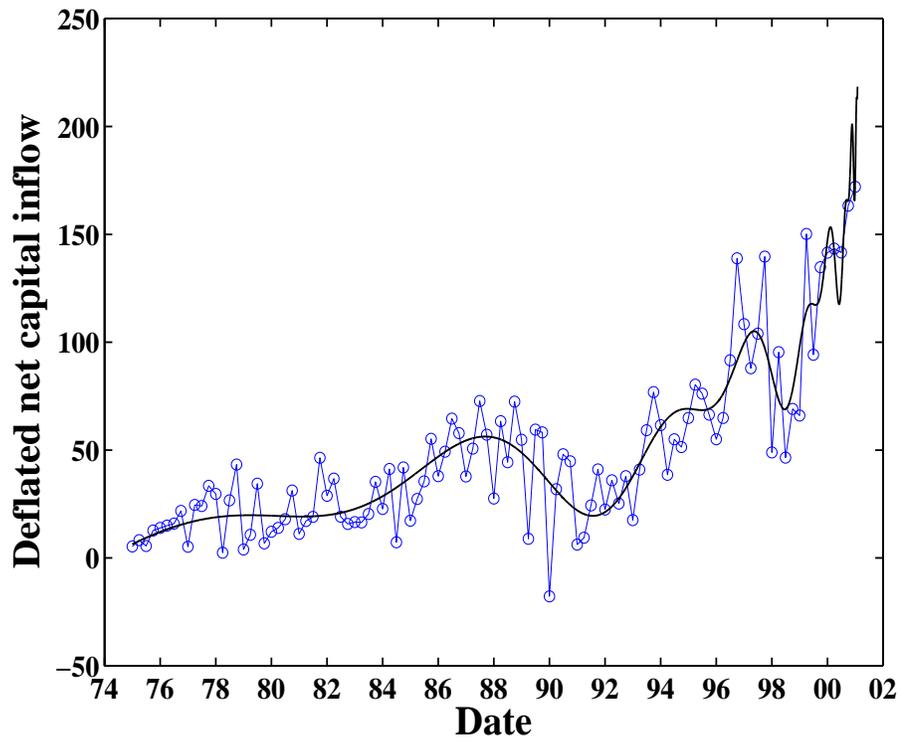,width=12cm, height=10cm}
\end{center}
\caption{Same as Fig.~\ref{Fig:BOPI} for the deflated
foreign net capital inflow in the USA. The predicted critical
time is $t_c={\rm{2001/03/14}}$, the power-law exponent is
$m=0.01$, and the angular log-frequency is $\omega=5.0$. The
fitted linear parameters are $A=3963$, $B=-3609$, $C_1=15.3$ and
$C_2=12.0$. The r.m.s. of the residuals of the fit is 16.114.}
\label{Fig:DeBOPI}
\end{figure}

We test for the robustness of this fit by studying the
impact of inflation on the extracted
log-periodic structures. We have deflated $I(t)$ for
inflation by the seasonally adjusted US consumer price index
(CPI), also available at the database FRED II. The original
monthly CPI were converted to quarterly CPI by averaging the three
monthly CPIs in each relevant quarter. Figure \ref{Fig:DeBOPI}
shows the time evolution of the deflated US foreign assets net
capital inflow $I(t)$ from 1975 to
the first quarter of 2001 and its fit to the second-order
Weierstrass-type function (\ref{Eq:It}). Again, four significant
log-periodic oscillations are observed in the figure, as well as
their second-harmonic oscillations, which exhibit very similar
characteristics as in Fig.~\ref{Fig:BOPI}. The predicted critical
time is $t_c={\rm{2001/03/14}}$, the power-law exponent is
$m=0.01$ constrained by the positivity condition,
and the angular log-frequency is $\omega=5.0$, which is essentially
undistinguishable from the value obtained without deflating the price.
The linear parameters are $A=3963$, $B=-3609$,
$C_1=15.3$ and $C_2=12.0$. The r.m.s. residual of the fit is
16.114. We conclude that inflation has a minor impact on the
LPPL structure as both $I(t)$ and the deflated
$I(t)$ exhibit the same accelerating oscillatory pattern, with
the key parameters, $\omega$ and $t_c$, which are very close.

\subsection{Robustness of the critical time $t_c$}
\label{s2:robust}

The determination of $t_c$ is particularly important as it gives
the estimated termination time of the speculative bubble as well
as the most probable time for a crash or a change of regimes.
Since the time evolution of a bubble is contaminated with noise
and the sparseness of the data close to the critical point, it
can be expected that the estimated $t_c$ may be quite sensitive to
the length and end-point $t_{\rm{last}}$ of the time series
used for its determination, as well as to the properties
of the noise \cite{bookCP,SJ01QF,bookcrash}. In our previous
studies, we often find that $t_c$
increases with $t_{\rm{last}}$, which makes the
prediction of $t_c$ unreliable. When the determination of $t_c$
is stable, this gives in general a reliable estimation of the
end of the bubble. In the present case, we find a remarkable
strong robustness of the prediction of $t_c$ when changing $t_{\rm{last}}$
both on the un-deflated $I(t)$ and the deflated $I(t)$.

The results of the test on un-deflated $I(t)$ are presented in
Table \ref{Tb1} with seven different values of $t_{\rm{last}}$.
For $t_{\rm{last}}$ varying from 1999/10/01 to 2001/01/01, the six
fits give almost the same solution with practically unchanging
values for the seven parameters: $t_c$, $m$, $\omega$, $A$, $B$,
$|C_1|$, and $|C_2|$. The predicted $t_c$ is found consistently in
the first quarter of 2001. When data after 1999/07/01 are excluded
( $t_{\rm{last}} \leq$ 1999/07/01), the solution changes suddenly,
with a new family of estimated $t_c=$ 1999/04/27 and a much
smaller value for the angular log-frequency $\omega=1.8$. Such low
value of $\omega$ can be interpreted as an artificial
log-frequency stemming from the most probable noise decorating a
power law \cite{HJLSS00}.

Table \ref{Tb1} thus shows that one could have predicted precisely
the crash date of the bubble on the net foreign inflow 1.5 year
before the crash happened!

\begin{table}[htb]
\begin{center}
\caption{\label{Tb1} Testing the robustness of fits with
expression (\ref{Eq:It}) of the seasonally adjusted quarterly
net capital inflow $I(t)$ not deflated for inflation. Notice the
robustness of the prediction of the critical time $t_c$ of the crash as well as
that of the angular log-frequency $\omega$ quantifying the log-periodicity.}
\medskip
\begin{tabular}{ccccccccccccc}
\hline\hline
$t_{\rm{last}}$&$t_c$&$m$&$\omega$&$A$&$B$&$|C_1|$&$|C_2|$&$\chi$\\\hline
2001/01/01 & 2001/03/12 & 0.01 & 4.9 & 7355 & -6719 & 21.5 & 16.2 & 22.810 \\
2000/10/01 & 2001/02/14 & 0.01 & 4.9 & 7243 & -6616 & 21.5 & 16.4 & 22.824 \\
2000/07/01 & 2001/02/06 & 0.01 & 4.8 & 7267 & -6639 & 21.9 & 16.3 & 22.882 \\
2000/04/01 & 2001/03/04 & 0.01 & 4.9 & 7304 & -6672 & 22.1 & 16.5 & 22.960 \\
2000/01/01 & 2001/02/05 & 0.01 & 4.8 & 7165 & -6545 & 22.0 & 16.3 & 23.056 \\
1999/10/01 & 2001/01/31 & 0.01 & 4.8 & 7161 & -6542 & 22.0 & 16.3 & 23.169 \\
1999/07/01 & 1999/04/27 & 0.01 & 1.8 & 2541 & -2277 & 40.5 & 22.4 & 21.956 \\
\hline\hline
\end{tabular}
\end{center}
\end{table}

Similar results for the deflated $I(t)$ are presented in Table
\ref{Tb2} for eight different values of $t_{\rm{last}}$. For
$t_{\rm{last}}$ varying from 1999/07/01 to 2001/01/01, the seven
fits give again almost the same solution with very close values of
the seven parameters. One can observe that the values of $t_c$ and
$\omega$ for the deflated $I(t)$ in Table \ref{Tb2} are almost the
same as for the un-deflated $I(t)$ in Table \ref{Tb1}. Notice,
however, the predicted $t_c$ using the un-deflated $I(t)$ are
closer to the actual time of the crash (in the foreign capital
inflow), compared with those using the deflated $I(t)$. This
suggests that the un-deflated $I(t)$ is a better measure for the
bubble as herding investors are more sensitive to the actual price
than their deflated value. The difference in values of the linear
parameters $A$, $B$, $|C_1|$ and $|C_2|$ in Table \ref{Tb2} and in
Table \ref{Tb1} simply reflect an affine transformation between
the two fitted second-order Weierstrass-type functions. For
$t_{\rm{last}} = 1999/04/01$ and for smaller values, the solution
branches to another family with incorrect critical times $t_c$.
These results show a rather amazing prediction power of
(\ref{Eq:It}) with respect to the determination of the critical
time.

\begin{table}[htb]
\begin{center}
\caption{\label{Tb2}Testing the robustness of fits and the
prediction of critical time on the seasonally adjusted quarterly
net capital inflow $I(t)$ deflated by seasonally adjusted consumer
price index for inflation.}
\medskip
\begin{tabular}{ccccccccccccc}
\hline\hline
$t_{\rm{last}}$&$t_c$&$m$&$\omega$&$A$&$B$&$|C_1|$&$|C_2|$&$\chi$\\\hline
2001/01/01 & 2001/03/14 & 0.01 & 5.0 & 3963 & -3609 & 15.3 & 12.0 & 16.114 \\
2000/10/01 & 2001/02/25 & 0.01 & 5.0 & 3924 & -3573 & 15.3 & 12.0 & 16.166 \\
2000/07/01 & 2001/02/14 & 0.01 & 4.9 & 3926 & -3576 & 15.5 & 11.9 & 16.231 \\
2000/04/01 & 2001/03/29 & 0.01 & 5.0 & 3945 & -3592 & 15.6 & 12.1 & 16.254 \\
2000/01/01 & 2001/04/12 & 0.01 & 5.1 & 3976 & -3621 & 15.6 & 12.2 & 16.332 \\
1999/10/01 & 2001/03/25 & 0.01 & 5.0 & 3951 & -3598 & 15.7 & 12.2 & 16.412 \\
1999/07/01 & 2001/06/12 & 0.01 & 5.2 & 3960 & -3606 & 15.1 & 12.6 & 16.394 \\
1999/04/01 & 1997/04/23 & 0.01 & 4.1 & 2101 & -1905 & 15.0 & 11.2 & 16.303 \\
\hline\hline
\end{tabular}
\end{center}
\end{table}

Additional tests on the goodness of the fits and on a non-parametric
analysis of the log-periodicity are offered in the Appendix. In particular,
we identify a rather strong periodicity in the residuals with a period
of exactly one year. Since the formula (\ref{Eq:It}) is describing
a scale invariance pattern unfolding as a function of the time $t_c-t$
towards a critical time, regular periodicities are just additional structures,
not relevant here. Our non-parametric analysis of the log-periodicity
gives a strong confirmation to our parametric fit (\ref{Eq:It}) and shows
that even higher harmonics of the fundamental angular log-periodic
frequency $\omega_1 \approx 5$ are present in the data. The
possible importance of harmonics in order to qualify
log-periodicity is made also more credible by recent analyses of
log-periodicity in hydrodynamic turbulence data \cite{turb1,turb2}
which have demonstrated the important role of higher harmonics in
the detection of log-periodicity.

\subsection{Comparison between the LPPL patterns in the NASDAQ 100 index and
the foreign capital inflow}

From the all-time high on the NASDAQ 100 Composite index of 5133
on March 10, 2000 to the low at 3321 on April 14, 2000, the NASDAQ
100 lost over 35\%. Johansen and Sornette have shown
\cite{Nasdaqcrash} that this crash was preceded by a speculative
bubble passing all LPPL tests, as defined in
Refs.~\cite{SJ01QF,bookcrash,endoexolppl}. The analogy with the
infamous crash of October 1929 was found to be striking: the
belief in what was coined a ``New Economy'' both in 1929 and in
the late 1990s made share-prices of companies with three digits
price-earning ratios soar.

Table \ref{tabnasdaq} presents a test of the robustness of the
fits with the second-order approximation of the Weierstrass-type
function (\ref{Eq:It}) of the NASDAQ 100 index, obtained by
varying the time interval of the analysis. We vary the start time
of the window of analysis, keeping the ending time fixed at the
date of March 10, 2000 (the date the all-time high was reached),
as in \cite{Nasdaqcrash}. The fits and predictions of $t_c$ are
very robust with respect to $t_{\rm{first}}$, confirming the
conclusions obtained previously \cite{Nasdaqcrash}.

\begin{table}[htb]
\begin{center}
\caption{\label{tabnasdaq} Tests of the robustness of the LPPL pattern on the
bubble that developed on the NASDAQ 100 Composite index in the late 1990s.
Each row corresponds to a different start time of the window
of analysis, keeping the ending
time fixed at the date of March 10, 2000. Each such time series is fitted
with the first-order approximation
of the Weierstrass-type function.
}
\medskip
\begin{tabular}{ccccccccccccc}
\hline\hline
$t_{\rm{first}}$&$t_c$&$m$&$\omega$&$\phi$&$A$&$B$&$C$&$\chi$\\\hline
1997/03/10&2000/05/25&0.18&7.41&0.08&10.50&-0.93&0.02&0.0609\\
1996/11/29&2000/06/17&0.01&7.89&0.29&48.88&-38.11&-0.06&0.0603\\
1996/08/22&2000/06/18&0.02&7.85&3.21&34.70&-23.99&0.05&0.0584\\
1996/05/14&2000/06/15&0.03&7.79&2.82&22.52&-11.96&0.05&0.0585\\
1996/02/05&2000/06/04&0.08&7.49&3.89&14.39&-4.21&-0.04&0.0574\\
1995/10/27&2000/05/17&0.15&6.95&3.36&10.99&-1.34&0.02&0.0574\\
1995/07/19&2000/05/08&0.18&6.68&4.73&10.31&-0.86&-0.02&0.0565\\
1995/04/10&2000/03/24&0.33&5.55&3.35&8.98&-0.18&-0.01&0.0613\\
1994/12/30&2000/05/06&1.00&0.11&4.01&9.09&-0.10&-0.10&0.0637\\
\hline\hline
\end{tabular}
\end{center}
\end{table}

In view of the joint detection of LPPL patterns in the NASDAQ 100
Composite index and in the net foreign capital inflow to the USA,
it is natural to ask whether the two are in any way related. If we
compare the three key parameters $m, \omega$ and $t_c$, we find
respectively: $m=0.01-0.18, \omega=7.0 \pm 1$ and $t_c=$ May 2000
for the NASDAQ 100 index and $m=0.01, \omega=5.0$ and $t_c=$ March
2001 for the foreign capital inflow. A smaller exponent $m$
implies a slower initial acceleration finishing more abruptly
close to $t_c$, hence suggests that the foreign capital inflow was
lagging behind the NASDAQ 100 index with respect to the unfolding
of the speculative bubble. The later date for $t_c$ found for the
foreign capital inflow suggests a confirmation of this lag.

\begin{figure}
\begin{center}
\epsfig{file=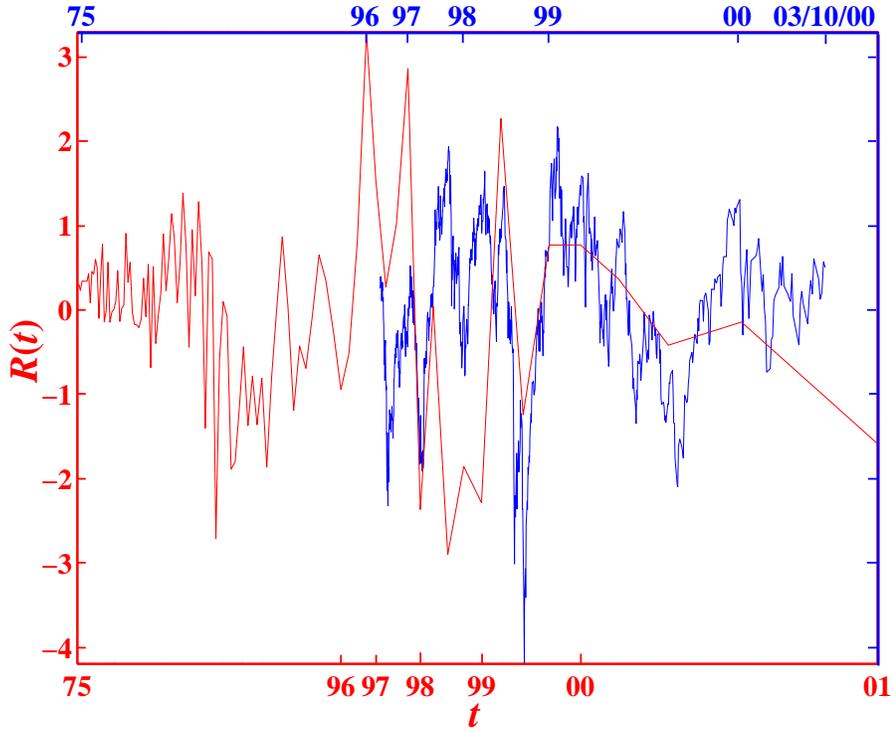,width=12cm, height=10cm}
\end{center}
\caption{This figure shows the two log-periodic residues $R_{Nasdaq}(t)$ and $R_I(t)$
defined as $R(t) \equiv [I(t) - A - B (t_c-t)^{m}]/(t_c-t)^{m}$ for the
the NASDAQ 100 index and for the foreign capital inflow. This construction extracts
the log-periodic components of the two time series,
while getting rid of their constant level, of
their trend and power law acceleration. These two log-periodic residues are
plotted as a function
of their respective $\ln (t_c-t)$ (recall that the two time series have
different critical times $t_c$). The thick (resp. thin) line and upper (resp. lower)
horizontal scale correspond to the NASDAQ 100 index (resp. foreign capital inflow).
}
\label{FigRLog2}
\end{figure}

In order to attempt to capture further the existence of a possible
synchronization between the two time series, we compare the pure
log-periodic structures within the two time series. For this, we
construct the log-periodic residues defined by $[I(t) - A - B
(t_c-t)^{m}]/(t_c-t)^{m}$ for the two time series, in order to
extract what should be pure log-periodic oscillations (if no other
pattern and noise were present). We then standardized these
residues to have unit variance, which provides us
with the standardized residues $R_{Nasdaq}(t)$ and $R_I(t)$.
Fig.~\ref{FigRLog2} shows $R_{Nasdaq}(t)$ and $R_I(t)$, plotted as
a function of their respective $\ln (t_c-t)$ (recall that the two
time series have different critical times $t_c$). The thick (resp.
thin) line and upper (resp. lower) horizontal scale correspond to
the NASDAQ 100 index (resp. foreign capital inflow). We note that
the last two log-periodic oscillations in the two time series
before the crash seem to exhibit a synchronization, {\bf when
taking into account a log-periodic lag}. Indeed, the local peak in
early 1999 in the NASDAQ 100 index residue $R_{Nasdaq}(t)$
corresponds approximately to the local peak in the foreign capital
inflow residue $R_I(t)$ in early 2000. Similarly, the local peak
in early 2000 in the NASDAQ 100 index residue $R_{Nasdaq}(t)$
corresponds to the local peak in the foreign capital inflow
residue $R_I(t)$ in mid 2001. Notice the two logarithmic scales in
terms of their respective $\ln (t_c-t)$ with different critical
times $t_c$ for each time series. This synchronization with a
log-periodic lag suggests that the foreign capital inflow has been
following a LPPL trajectory similar to that of the NASDAQ 100
index, indicating a speculative bubble. However, this signal on
the foreign capital inflow has been lagging in a {\bf
self-similar} way behind the NASDAQ 100 index bubble. The
reference to a log-periodic lag stresses that this lag was not
constant in time but has been shrinking as the bubble approached
its termination. The synchronization is however only approximate
as the angular log-frequencies are different, indicating a
transient effect present only at the end of the bubble. The fact
that parallel log-periodicities can culminate at different
critical times is made possible by the extended description in
terms of a hierarchy of singularities offered by the
quasi-Weierstrass functions \cite{GS02PRE}.

\section{``Egg-and-Chicken'' source of the NASDAQ 100 bubble
and of the net capital foreign inflow bubble to the USA \label{jelle}}

The evidence presented in the previous section suggests that
foreign investors have been attracted by the capital gains offered
by the bubble on the US stock market. They have followed the
bubble rather than having been at its origin. The synchronization
of the log-periodic patterns of the NASDAQ 100 index and of the
foreign capital inflow however indicates that the foreign capital
inflow may have amplified the bubble, in line with the standard
scenario $1-5$ of a bubble described in the introduction.

We now attempt to elaborate a bigger picture embodying what could
be the major elements linking foreign capital inflows and the
stock market bubble. We also try to understand the positive
feedbacks between the two processes and how they may be linked
with other macroeconomic processes.

\begin{figure}
\begin{center}
\epsfig{file=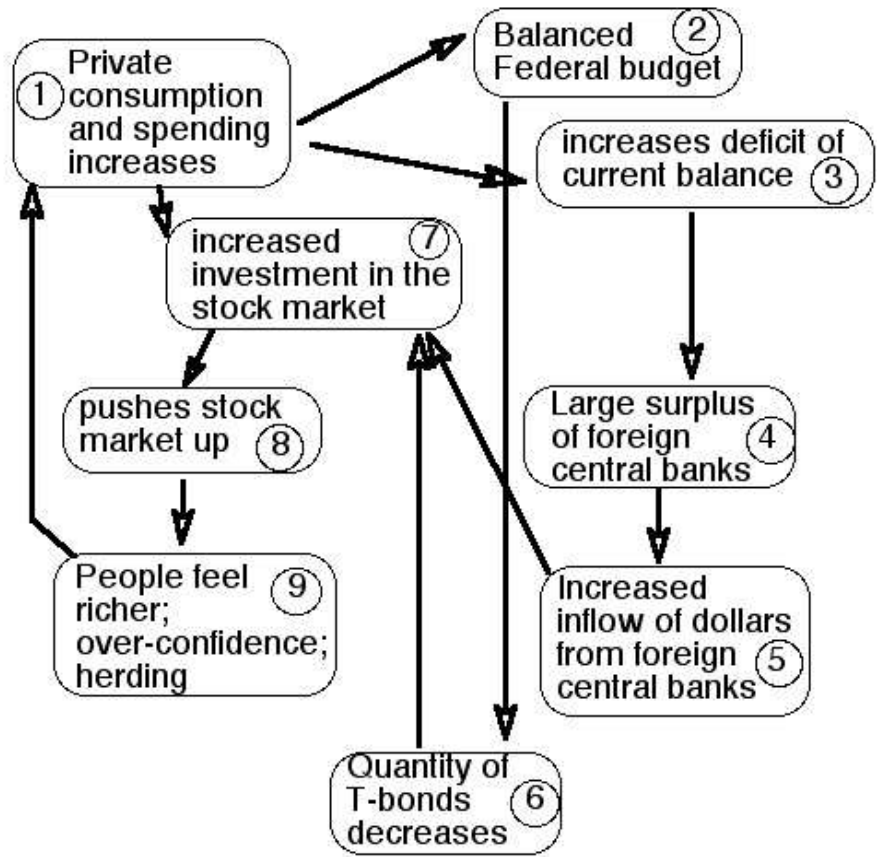,width=12cm}
\end{center}
\caption{Flowchart of the feedback loops between
market-appreciation / increased-spending /
increased-deficit-of-balance-of-payment / larger-foreign-surplus /
increased-foreign-capital-inflows, which are proposed to explain
the speculative bubble in the US stock markets and to rationalize
the synchronization of the LPPL patterns shown in
Fig.~\ref{FigRLog2}. According to this scenario, the 1990s
expansion in the US was powered in large part by a large increase
in net spending by the private sector, which itself was fueled by
the exhuberance and optimism resulting from the feedback of the
growing stock market via the wealth effect on spending.}
\label{Macrolinks}
\end{figure}

We start from the premise that the main mechanism behind the
development of speculative bubbles is the strengthening of herding
and over-confidence among investors, providing positive feedback
which further enhances the expectation of future capital gains
\cite{Galbraith,Garber,emerg,bookcrash}. In the case of the
internet and information bubble that ended in 2000, the flowchart
shown in Fig.~\ref{Macrolinks} summarizes the main processes
involved and illustrate the cause and effects of the foreign
capital inflow.
\begin{itemize}
\item[(i)] Box 1 in Fig.~\ref{Macrolinks}: It is a well-documented
fact that the private disposable income over expenditure has gone
from its long-term average surplus of about 3\% GDP per year in
the early 1990s to a deficit approaching 6\% GDP in 2000
\cite{Godley}. By ``private'', we refer to the aggregate of
households and corporation.

\item[(ii)] By accounting balance, the surplus of private
disposable income over expenditure is equal to the government
balance (written as deficit) plus the current balance of payment
(income from exports minus expenditure to pay for imports):
Private surplus $=$ Government deficit $+$ balance of payment.

\item[(iii)] Boxes 2 and 3 in Fig.~\ref{Macrolinks}: By (ii), the
large private deficit has fostered a growing deficit of the
balance of payment and a diminishing deficit of the Government
turning to surplus in 2000. The larger spending of the private
sector has profited to foreign manufacturers who have been
increasingly successful in invading U.S. markets. The
strengthening of the US dollar has also helped foreigners exports
to the US. The growing deficit of the balance of payment has also
resulted from the increased outsourcing of intermediate products.

\item[(iv)] Box 4 in Fig.~\ref{Macrolinks}: The annual deficit of
the balance of payment has been growing and reached \$500 billion
(2\% GDP). This deficit corresponds mechanically to a surplus for
foreign exporting nations, which have accumulated trillions of
dollars in reserves since the early 1970s. In 2001, the United
States' net debt to the rest of the world jumped to \$2.3 trillion
(this amounts to about 25\% of GDP), a level double that recorded
in 1999 \cite{Tille}. Much of the increase reflects new borrowing
undertaken by the country to finance its mounting current account
deficit. A third of the change, however, could result from a
simple accounting effect, the impact of a rising dollar on the
decreasing value of U.S. assets held abroad.

\item[(v)] Box 5 in Fig.~\ref{Macrolinks}: To avoid pushing up
their own currencies in order to keep their competitive hedge and
because of the large interest rates enjoyed until 2000 on US
treasury bonds, foreign central banks reinvested a significant
part of their reserve denominated in US\$ back in the US, leading
to a large inflow of capital (see Figs.\ref{Fig:It} and
\ref{Fig:BOPI}). Thus, the spent US\$ fueling the deficit of the
balance of payment have been coming back as a boomerang in the
form of investments by foreign countries \cite{Duncan}.

\item[(vi)] Boxes 6 and 7 in Fig.~\ref{Macrolinks}: However, due
to the decreasing deficit of the Federal government, a decreasing
quantify of treasury bonds are been issued, letting the stock and
real-estate markets (Fannie Mae, corporate bonds, NASDAQ 100) as
the main siphoning tanks of the foreign capital inflows.

\item[(vii)] Boxes 1 and 8 in Fig.~\ref{Macrolinks}: This inflow
of foreign capital together with an exhuberant herding mood on the
part of investors \cite{Shillerexu} has fueled the bubble in the
later part of the 1990s.

\item[(viii)] Boxes 9 and 1 in Fig.~\ref{Macrolinks}: As a result
of the impressive appreciation of the stock market, investors and
households have felt richer, spending and consuming wantonly,
further fueling the bubble by the feedback of this spending on
economic activity, leading to a positive feedback amplification.
This is the so-called wealth effect such that an increase in
wealth directly causes households to increase their consumption
and decrease their saving (see for instance \cite{Makiwealth} and
references therein).

\end{itemize}

This implies that the 1990s expansion was powered in large part by
a large increase in net spending by the private sector, which
itself was fueled by the exhuberance and optimism resulting from
the feedback of the growing stock market via the wealth effect on
spending. This also suggests a mechanism for the observed
synchronization of the foreign capital inflow documented in this
paper and the stock market LPPL pattern. First, the LPPL pattern
of the foreign capital inflow documented here shows that foreign
investors were following a herding behavior similar to that of
national investors. Second, the feedback loops between
market-appreciation / increased-spending /
increased-deficit-of-balance-of-payment / larger-foreign-surplus /
increased-foreign-capital-inflows rationalize our finding in
Fig.~\ref{FigRLog2} of a synchronization of the LPPL pattern taken
as the qualifying signature of a speculative herding bubble.

Another interesting question is the following (we are indebted to
J. Spanos for these remarks): why did it take so long (well into
2001) for the foreign capital to stop flowing into the US, while
the US markets have crashed almost a year earlier? It seems that
foreigners did not acknowledge the coming bear market in stocks
until it was well under way. This implies that the foreigners did
not cause the crash of the stock market. Their actions simply
confirmed it after the fact, with a considerable lag. This
confirmation came well before the 2001 attacks on the world trade
center on 911. This is a strong evidence in proving that the
attacks had nothing to do with the bear market which was already
on its way and confirmed by foreign inflows. We also note that
both the NASDAQ crash time and the foreign inflows crash time
could have been forecasted as early as 1999.

\section{Implications for the following years}

Let us examine the implications of our findings and reasoning for the next years.

Two main variables have changed since 2000, when compared with the situation
in the late 1990s described in the previous section \ref{jelle}.

First, the stock markets have declined substantially since their
all-time highs in the first quarter of 2000. The NASDAQ 100 has
lost close to 75\% of its value at its dip in October 2002, and
remains at a loss of close to 70\% at the time of writing. The
S\&P500 has lost about 47\% at its dip in October 2002 and remains
negative by about 35\%. The Dow Jones Industrial Average lost
about 38\% at its minimum in October 2002 and remains off by about
23\%. The wealth effect has thus vanished.

Second, the Federal budget surplus of the last years of the 1990s
has transformed again into a growing deficit. The private sector
spending fury has abated but the private disposable income over
expenditure remains in deficit and several percent of GDP below
its long term average \cite{Godley03}. A radical change of
attitude to budget deficits has also occurred, which suddenly
became respectable as a way to fight fears of recession
\cite{fearrec}. In addition, after 911, private spending of
households were encouraged at the highest level of the executive
hierarchy as being patriotic.

Our own analysis for the period from 2000 to 2002 strongly suggest
that the US stock markets has declined as a consequence of another
herding process characteristic of bearish ``anti-bubbles''
\cite{SZ02QF,ZS02,ZS03}.

The overheating of the speculative bubble led to a crash on the
NASDAQ 100 in March-April 2000. As a consequence of vanishing of
the wealth effect and of the self-reinforcing negative sentiments,
the stock markets have gone down. As a result of the increasing
Federal budget deficit, foreign capital have flowed again to buy
the debts issued by the treasury, disrupting a part of the flow
that was previously directed to the stock markets. As
Fig.~\ref{Fig:It} and our analysis show, the crash on the foreign
capital inflow coincide with the transition from surplus to
deficit in the Federal budget.

There are several drastically different scenarios, proposed by
various analysts and commentators, for the future evolution of the
stock market and of the economy. Standard measures of valuations
suggest that the market has not yet bottomed as it still appears
significantly over-valued; currently, the price-to-earnings ratio
is over 34, the dividend yield is 1.74\%, and the price-to-book
value is over three times. Compare this to the bottom in 1982, for
which the price-to-earnings ratio of the S\&P 500 was 7, the
dividend yield was 6.3\%, and the index was selling at book value.

\begin{figure}
\begin{center}
\epsfig{file=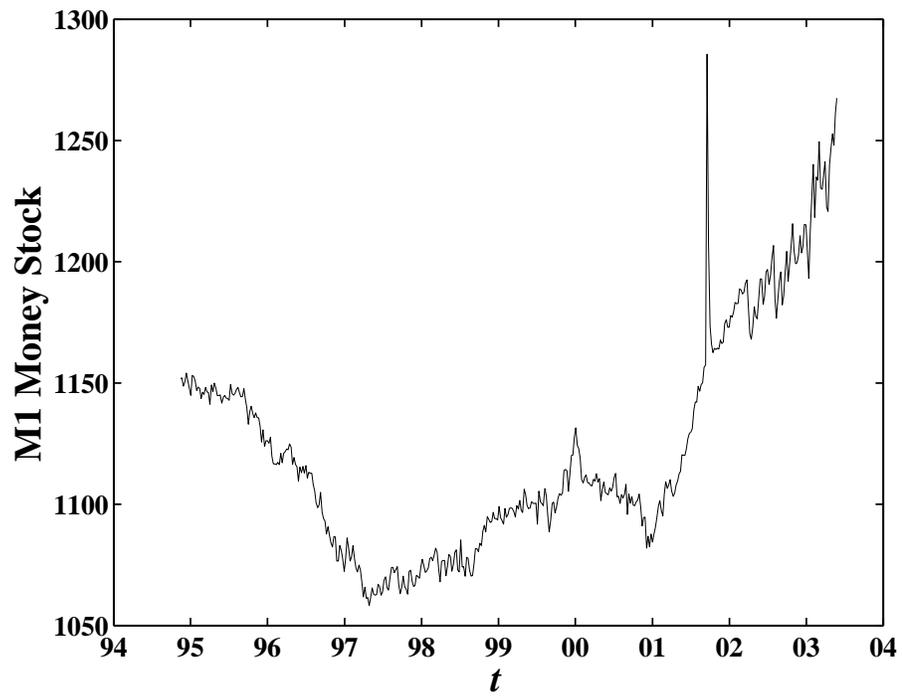,width=12cm}
\end{center}
\caption{Growth of the money supply M1, which is basically cash
and checking accounts. The outlying peak is the response to the 911 attack.} \label{M1}
\end{figure}

However, there are strong forces willing to push up the confidence
of investors, to foster the economy and by the same token the
stock market, since the later is a confidence/sentiment
thermometer. A first force is the increase in money supply. Since
early 2001, the money supply M1, which is basically cash and
checking accounts, has been rising at a 30\% annual rate, with a
deceleration in 2002 and then resuming an acceleration in the
first part of 2003, as seen in Fig.~\ref{M1}. The size and growth
of money supply is influenced (controlled?) by the Federal Reserve
System through its direct control over the reserves of member
banks, the discount rate and through open market operations. This
increasing money supply, which is supposed to foster economic
development, also finds its way in the stock market, because
companies are not spending the money to boost capacity, as some
industries like semiconductors are working at only 65\% capacity
utilization, and the overall capacity utilization rates are around
75\% (compared to +90\% in the late 1990s). Companies are also
using this cheap money to re-leverage their balance sheets,
similar to consumers switching their credit card debts to
different cards with lower rates.

A second force is found in the behavior of foreign capital
inflows. With the growing availability of treasury debts, the
enormous surplus of foreign central banks have found again a
natural depository, which avoids the risk of inflating their own
currency. However, the interests paid have now plummeted from
above 6\% to slightly above 1\% per year. The bonds are attractive
only on the basis of a speculative capital appreciation, no more
by the paid interests. Therefore, naturally, foreign capital is
attracted to the US stock markets. And here comes into play a
confidence and herding game. Notwithstanding the stock market
over-value, foreign capital (as well as national investors) would
like to see the stock market re-appreciate since they have not
many other choices to invest their surplus dollars. There is thus
a growing availability of capital to hold prices from falling, at
least for a while.

This may constitute a part of the explanation for the appreciation
of the US stock market since the uncertainties with the war with
Iraq in March 2003 fadded out. Will this continue? At this stage,
since we view investors confidence and herding as an important and
integral part of the self-organization of stock markets and of the
economy, it is interesting to dwell more on quantitative measures
of confidence. So-called market sentiment ratings are obtained
through polls where responses are bullish, bearish or neutral on
the market, which are regularly available \cite{Barron}. During
the entire time when the NASDAQ 100 dropped by more than 75\%
since March 2000, there was not one weekly reading showing more
bears than bulls. It took the greatest terrorist act in U.S.
history to finally register a week during which there were more
bears than bulls, for the first time in 153 weeks.  The plurality
of bears over bulls was minuscule considering the magnitude of the
events. The 911 catastrophe could only produce three consecutive
weeks where bears outnumbered bulls, after which the bulls
dominated the sentiment readings again. The study reported in
Barron's study on May 5th, 2003 gave the following categories and
responses:  Very Bullish (9\%); Bullish (51.1\%); Neutral
(28.6\%); Bearish (10.5\%); and Very Bearish (0.8\%). This is not
the type of sentiment (60.1\% bulls versus 11.3\% bears) after
three years of a nasty bear market! This overwhelming reading,
already impressive in a bull market, is unheard of for a bear
market.

Our reading of this surprising pervasive bullish sentiment is that
it confirms that the private and foreign investors {\bf want} the
market to go up, but that there is so much uncertainty that
``wishing'' is different from ``acting,'' that is, investing.
Investors are waiting for signs of confirmation of their bullish
sentiment to drive the price up. They have already be burned
severely by the crash in 2000 and the two years that followed. In
particular, foreign investors have strong incentive to buy the US
stock market as well as corporate bonds and the debts in US real
estate market \cite{house} in order to get a return on their
surplus dollars above the ridiculous discount rate offered on
treasury bonds. But such action would be warranted only if the
market risk is not too high, hence the conflicted observations of
a strong bullish sentiment in a depreciating stock market. The
contradictory conclusions on the economic outlook and these polls
suggest that the natural herding behavior of investors will be
even more predominant in the future and lead to highly volatile
and unstable market behavior in the near future.

As the US stock market and economy were heating up in the late
1990s, the higher interest rates and stronger dollar were the
natural instruments to attempt to avoid inflation and to try to
stabilize growth but also resulted from the economic and stock
market growth \cite{VLAARD}. Actually, as we explained above, both
led to the rather perverse effect of fueling further the bubble by
(1) increasing the deficit of the balance of payment through the
deterioration of competitiveness accompanying a strong dollar and
(2) by the attractiveness of investing in the US for foreign
capital in part obtained through the surplus of foreign countries
on their balance of payment.

Now, the situation is different. The economy has been flirting
with stagnation and depression several times in the last two years
and the stock markets have been falling down; hence, the massive
cuts in interest rates. Since May 2002, the strong dollar has been
steadily losing ground against the major currencies, with an
acceleration of this loss since November 2002, showing a
cumulative loss of about 28\%. As a consequence, investments in
the US by foreigners is becoming less attractive due to the
increasing exchange risk and the lowest interest payment.

Weighting these different ingredients, our prefered scenario for
the future is the following.
\begin{itemize}
\item The private sector will continue spending more than its
long-term average, as it is psychologically difficult to abandon
habit acquired in good times (the glorious 1990s) and it is in
addition almost considered as a patriotic act.

\item The debt of the Federal Government as well as the private,
municipal, corporate and local government debts will continue to
rise, reinforcing further the US as the major deficit nation.

\item As a consequence, interests will remain low to allow
servicing of the payment of the interests of the debts, both of
private sector and the government. This will continue to have the
effect of further fueling the growth of liquidity by the mechanism
of fostering loans refinancing on lower interest rates (mostly
from residential real estate)

This consequence will also continue to be a source of the two
first bullets, acting as a positive feedback loop. The central
bank of the US is now compelled to peg short-term interest rates,
promise to forewarn the marketplace of any intention to adjust the
peg. and to guarantee continuous marketplace liquidity. Federal
Reserve operations will continue to work by aggressively
manipulating rates, yield spreads (by repurchasing long-term
debts) and, increasingly, market perceptions to ensure these
goals. This is further reinforced by the perception that the
proposed tax cuts and current low interest rate environment will
further increase liquidity and turn the US economy around which
will power the stock market even further ahead.

\item The dollar will continue its descent as a mechanism to fight
against the deficit of the current balance by boosting exports
(which translate into cheaper imports for foreigners). A decrease
of the dollar also provides a mechanical device to decrease the
absolute value of the debt. This will accentuate the incentive for
foreign central banks to sell progressively their dollars, but
they cannot do it too fast to avoid losing the competitiveness of
their currencies. There is thus a subtle balance between the
economic competition giving rise to surplus in dollars, the
corresponding importance of not having a strong currency and the
present lack of attractiveness of the dollar. We thus envision a
slow sell out of the US dollar, but only on a limited scale since
the world is overflowed by dollars, which has replaced gold at the
international reserve currency, since the breakdown of Bretton
Woods (about 80\% of the world's free capital is invested in
dollars, even though the US makes up only about 30\% of the
world's economy).

\item In view of these negative factors, foreign capital will be
less attracted to the stock market.
\end{itemize}

\begin{figure}
\begin{center}
\epsfig{file=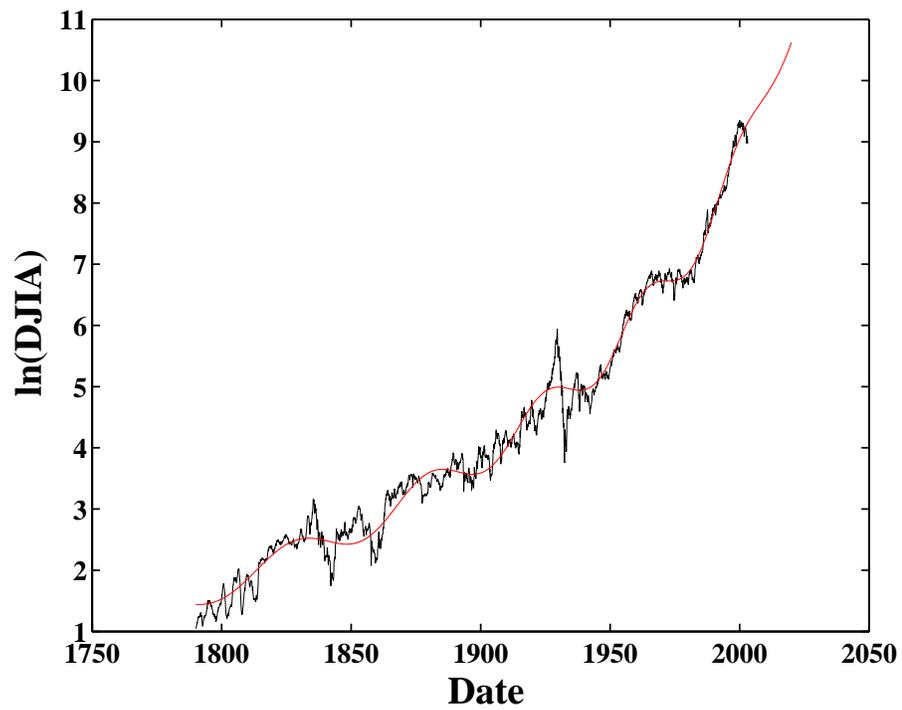,width=12cm}
\end{center}
\caption{Long term view of the evolution of the US market proxied
by the DJIA and its extrapolation in the past as explained in
\cite{twothousandfifty}. The continuous line shows the fit with
our LPPL formula extended to describe such long time intervals.
See \cite{twothousandfifty} and Chapter 10 of \cite{bookcrash} for
details. } \label{FigDJIAfit}
\end{figure}

\begin{figure}
\begin{center}
\epsfig{file=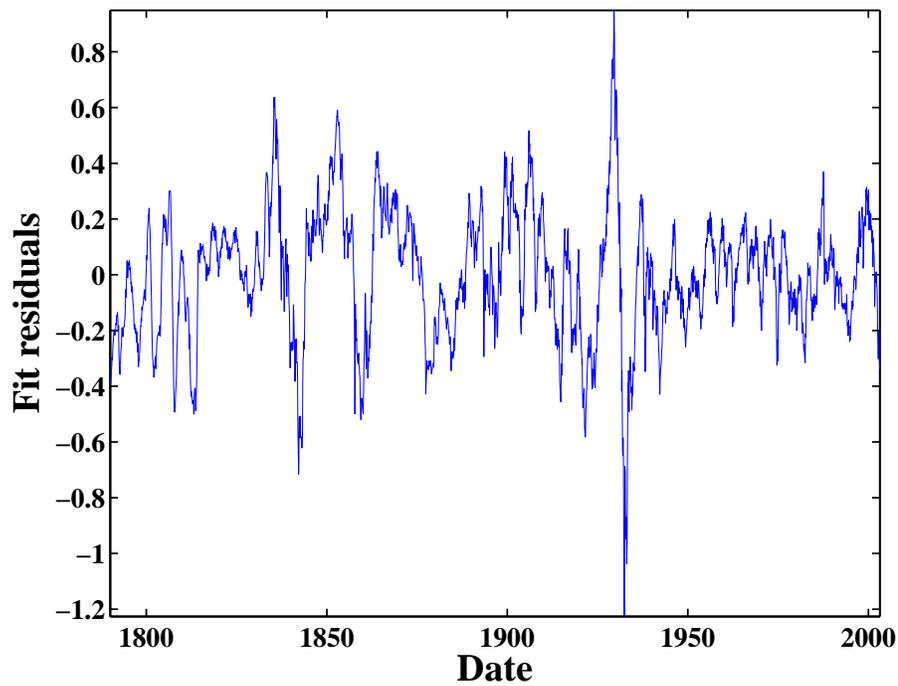,width=12cm}
\end{center}
\caption{Residual or difference between the realized DJIA and the
fitted formula shown in Fig.~\ref{FigDJIAfit}. The previous behaviors
of the residuals and the characteristic time scales before
recovery suggest that one may wait for a year or two before the
stock market recovers its long-term trend. } \label{FigDJIARes}
\end{figure}

From the point of view of detecting large scale cooperative
behavior, Fig.~\ref{FigDJIAfit} presents a long term view of the
evolution of the US market proxied by the DJIA and its
extrapolation in the past as explained in \cite{twothousandfifty}.
The continuous line shows the fit with our LPPL formula,
a second-order Landau expansion described in \cite{twothousandfifty,bookcrash},
which is an extension of the formula (\ref{Eq:It}) to
describe such long time intervals. Fig.~\ref{FigDJIARes} shows the
residual or difference between the realized DJIA and the fitted
formula. The previous value of the residuals and the
characteristic time scales before recovery suggest that one may
wait for a year or two before the stock market recovers
its long-term trend. This
appears to be in line with the prediction that the US stock market
will bottom during or at the end of the first semester of 2004,
according to our previous analysis of the LPPL anti-bubble which
started in 2000 \cite{SZ02QF}.

The residuals plotted in Figure \ref{FigDJIARes} show that the
current bear market has been almost as severe as all other bear
markets since 1950 but far less severe than the 1932 bear market.
One scenario is that the market may have in fact bottomed out in
October of 2002. However, if this is a deflationary bear market as
the US was in 1932, then a huge decline is still ahead. The
interest rate data so far point to a ``deflationary'' environment
but this is far from conclusive as the Federal Reserve is
currently waging war with deflation and is prepared to drive short
term rates to zero to avoid this scenario. But doing so, it is
fueling the credit bubble to unprecedented levels, developing
another dimension of instability.

\begin{figure}
\begin{center}
\epsfig{file=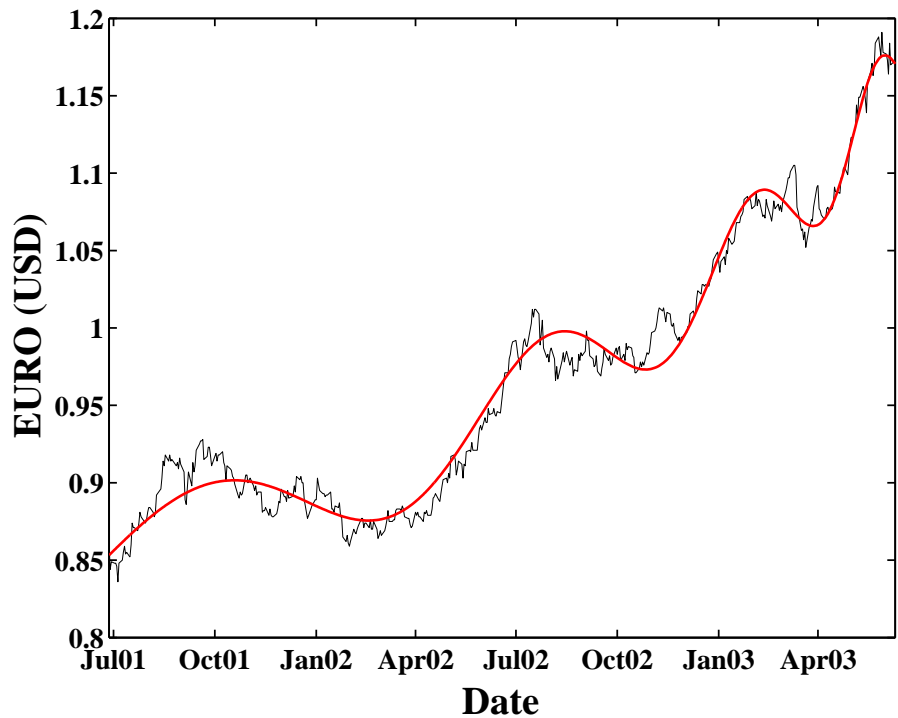,width=12cm}
\end{center}
\caption{Time evolution of the EURO in US\$ (thin fluctuated line)
and its LPPL fit (thick smooth line). The fitted
critical time is $t_c = \rm{2003/10/30}$ with a power-law exponent
$m = 0.13$ and angular log-frequency $\omega = 12.0$. The r.m.s.
of the fit residuals is $\chi = 0.0125$. The log-periodicity
is highly significant with almost four oscillations.} \label{FigEURO2USD}
\end{figure}

\begin{figure}
\begin{center}
\epsfig{file=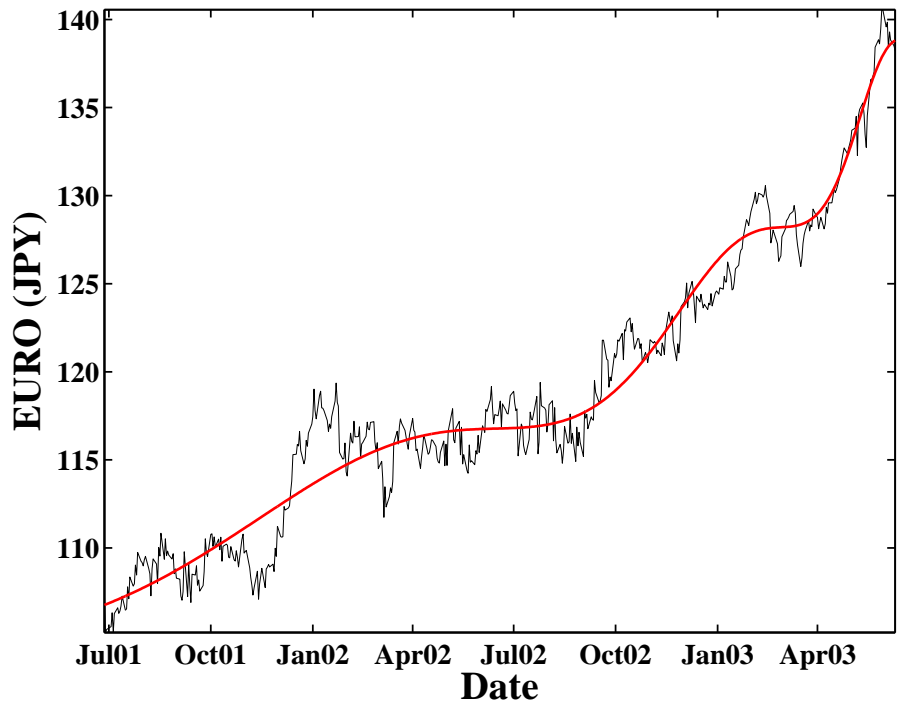,width=12cm}
\end{center}
\caption{Time evolution of the EURO in Japanese Yen (thin fluctuated line)
and its LPPL fit (thick smooth line). The fitted
critical time is $t_c = \rm{2003/09/04}$ with a power-law exponent
$m = 0.09$ and angular log-frequency $\omega = 7.3$. The r.m.s. of
the fit residuals is $\chi = 1.635$. } \label{FigEURO2JPY}
\end{figure}

\begin{figure}
\begin{center}
\epsfig{file=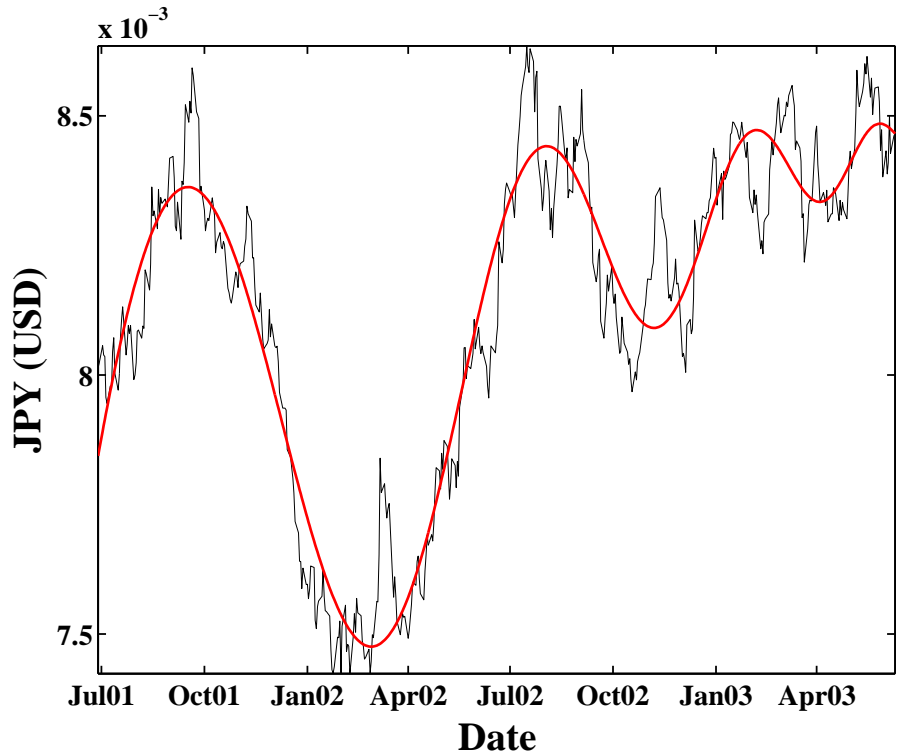,width=12cm}
\end{center}
\caption{Time evolution of the Yen in US\$ (thin fluctuated line).
The thick line is its fit with the LPPL formula (\ref{Eq:It}) with $n=1$
(a similar result is obtained with $n=2$). The parameters of the fit are
$t_c = \rm{2003/10/30}$,
$m = 1.74$ and angular log-frequency $\omega = 11.8$. The r.m.s.
of the fit residuals is $\chi = 9.78\times 10^{-5}$. The value larger than $1$ of the
exponent $m$ implies the absence of any acceleration of the appreciation of the 
Yen in US\$. }
\label{FigJPY2USD}
\end{figure}

A very interesting additional information is provided by the
behavior of the main currencies against the US dollar. We have
found unmistakable LPPL signatures of a speculative bubble which is
presently developing on the EURO. Specifically, Figure
\ref{FigEURO2USD} shows the EURO in US\$ and a typical 
accelerating LPPL bubble
pattern, which is suggestive of a speculative herding buying of
EURO's using US\$. Figure
\ref{FigEURO2JPY} shows the EURO in Yen and again a significant 
accelerating LPPL pattern.
In contrast, the Yen in US\$ does not have any acceleration
(nor has the US\$ in Yen), even if 
a marginally significant log-periodicity may be observed, as shown in 
Figure \ref{FigJPY2USD}. These three figures provide a remarkable message:
the depreciation of the US\$ is not just the undirected flight-for-safety of a herd
fleeing from a looming catastrophe; it
seems to be associated with a speculative bubble directed to what is felt (at least on the 
short- and medium-term) to be the new haven currency, the EURO.

\section{Conclusion}

Our main conclusions are the following.
\begin{enumerate}
\item The ``sacrifice'' of the US\$ and the stock market is the
cost for a lower sustainable debt burden on the global US debt
structure and for allowing the slow rebuilding of an
internationally competitive economy. This is reinforced by the
evidence for a speculative bubble developing on the EURO. We thus
envision a continuation of the depreciation of the US\$ that may
reach unprecedented low levels.

\item On the medium term, the stock market is not going to recover
a strong bullish trend as the growing Federal deficit is consuming
a large part of the foreign surplus dollars and the stock market
is remaining a very risky and unattractive investment. In
addition, the huge credit bubble, that the US has developed in the
last decade and is increasingly fueled in the hope ``to avoid
deflation,'' may be expected to burst and have severe consequences
for the recovery of the economy.

\item On the short term, the stock market may hold for a while as
one of the main sink of a strong surge of liquidity and of the
credit bubble, justified in the mind of investors by their
sentiment and hopes.
\end{enumerate}

The US has been growing as the major deficit nation in the world, 
attracting huge
amounts of foreign capital. In parallel, it is also growing
steadily by immigration, powered by a variety of factors. Studying
the relationship between immigration and capital flows, Groznik
has shown that, surprisingly, labor not only moves in the same
direction as capital, but it also leads capital \cite{Groznik}.
This finding is also found for various countries, periods and
migration flow specifications. Thus, an important predictive
variable for international capital flows is immigration flows.
One strength of the USA has been its ability to attract people
and capital. To what degree this inflow will continue to justify its
unsustainable deficits (at all levels) is linked with its potential
for development of new riches and remain to be seen.

\textbf{Acknowledgments}

We are grateful to Gary Dostourian who has attracted our attention
to the foreign capital inflows, and to Didier Darcet, Garrett
Jones, David Nichols and John T. Spanos for stimulating
discussions. This work was supported in part by the James S. Mc
Donnell Foundation 21st century scientist award/studying complex
system.

\section{Appendix: goodness of fit and non-parametric analysis of log-periodicity}

\subsection{Testing the goodness of the fits}
\label{s2:goodness}

We present two tests assessing the goodness-of-fit of the
seasonally adjusted quarterly US net capital
inflow $I(t)$ of foreign assets with formula (\ref{Eq:It}).
They are based on the analysis of the
fit residuals with the autocorrelation method
\cite{FC92CJP} and the sign test \cite{M01JMS}.
Based on a rigorous mathematical proof \cite{F03JMS},
these two tests are
complementary in their detection of possible remaining
patterns in the residuals. The autocorrelation method is often
considered to
provide an estimate of the model error. The sign test
is based on an intuition that the changes of sign of the residuals
should be random for a good fit.

Assume that $\{y_i: i = 1, \cdots, n\}$ are $n$ observations of
the true values $\phi_i$ at $n$ evenly spaced successive times
$t_i$ and $f(t; \Theta)$ is a proposed model with $p$ parameters
$\Theta$. If $\hat{\Theta}$ is an estimate of $\Theta$, then the
residuals of the fit are
\begin{equation}
 e_i = y_i - f(t_i; \hat{\Theta})~.
 \label{Eq:ei}
\end{equation}
The one-lag covariance of the residuals is then defined
\begin{equation}
 a = \frac{1}{n-1}\sum_{i=1}^{n-1}e_i\times e_{i+1}~,
 \label{Eq:a}
\end{equation}
where the mean of $e_i$ is zero (this has been verified to be true
in our residuals with extremely good accuracy $\sim 10^{-11}$).
The unbiased estimate of the variance of the residuals is
computed as
\begin{equation}
 s^2 = \frac{1}{n-p}\sum_{i=1}^{n}e_i^2~.
 \label{Eq:s2}
\end{equation}
By assuming that (1) the fit is made using least-squares, (2) the
measurement errors $\epsilon_i=y_i-\phi_i$ are centered random
variables with the same variance, and (3) the data are
uncorrelated, one can deduce the first order estimates for the
model error $\sigma_\delta$ or model bias, defined as the standard deviation
of $\phi_i-f(t_i;\hat{\Theta})$, and the measurement error $\sigma$,
defined as the standard deviation of $\epsilon_i$:
\begin{equation}
\begin{array}{lll}
 \sigma^2 & \equiv & {\rm Var}[y_i-\phi_i] \approx (1-\frac{p}{n})s^2-a~,\\
 \sigma_{\delta}^2 & \equiv & {\rm Var}[\phi_i-f(t_i;\hat{\Theta})]
 \approx \frac{p}{n} s^2 + (1-\frac{p}{n})a~.
 \label{Eq:sigma}
\end{array}
\end{equation}
For $p/n=0$, Eq.~(\ref{Eq:sigma}) recovers the standard
zero order estimations. The key statistics of the sign
test is the frequency $f$ of changes of signs of the residuals
$e(t_i)$ given by \cite{F03JMS}
\begin{equation}
 f \approx \frac{1}{2}-\frac{a}{\pi s^2}~.
 \label{Eq:f}
\end{equation}

\begin{figure}
\begin{center}
\epsfig{file=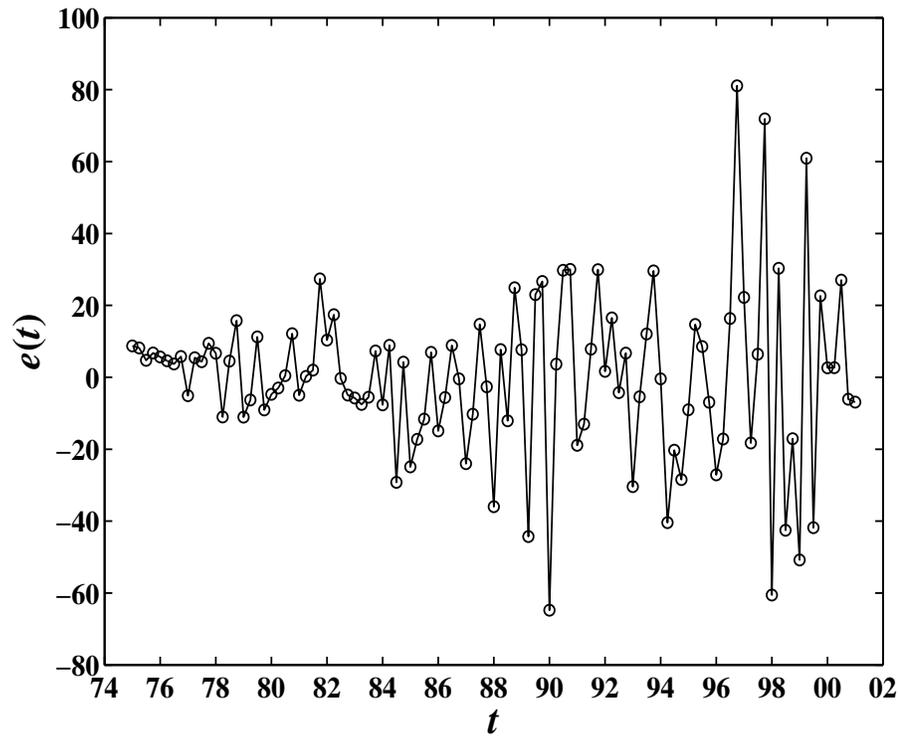,width=12cm, height=10cm}
\end{center}
\caption{Time evolution of the residuals $e(t)$ of the fit shown
in Fig.~\ref{Fig:BOPI}. The one-lag correlation coefficient is
$-0.16$ showing anti-persistence. The frequency of change of signs
of the residuals is ${\it f}=0.55$.} \label{Fig:GOF}
\end{figure}

\begin{figure}
\begin{center}
\epsfig{file=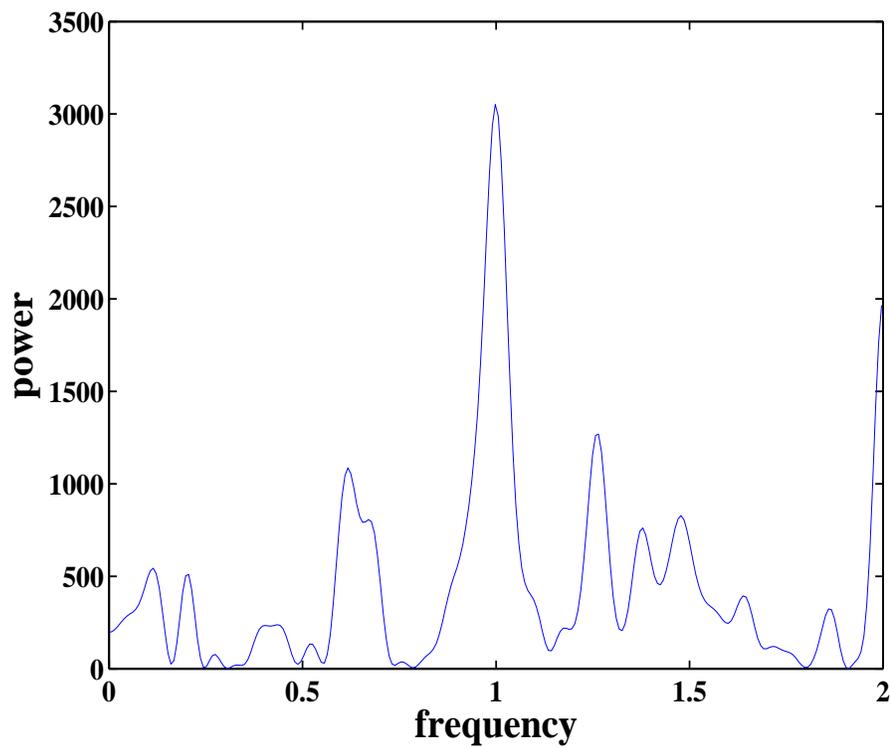,width=12cm, height=10cm}
\end{center}
\caption{Spectrum of the residuals $e(t)$ shown in
Fig.~\ref{Fig:GOF}. The very significant peak corresponds to a
periodicity of one year.} \label{specres}
\end{figure}

The residuals of our fit are shown in Fig.~\ref{Fig:GOF}.
We have $p=5$ free parameters since the four
linear parameters are slaved \cite{ZS03}. We find $a = -86$ and
$s^2 = 546$, which gives a one-lag correlation coefficient $a/s^2 = -0.16$.
This yields $\sigma^2 = 602$,
$\sigma_\delta^2 = -56$. The negative value of $\sigma_\delta^2$
reflects some anti-persistence in the residuals. The frequency of changes of
signs given by (\ref{Eq:f}) is
${\it{f}} = 0.550$, which is not too far from the value $0.5$ for
completely random residuals. We cannot however negate the existence of
a significant residual structure, whose prominent characteristic is a
strong yearly periodicity, as shown from the spectrum of the residuals
presented Fig.~\ref{specres}.

\subsection{Generalized $q$-analysis of the log-periodic
structure} \label{s2:sig}

We check the
the significance level of the extracted log-periodic
pattern using a non-parametric analysis, called the
$(H,q)$-analysis which we have already explained and used in our previous works
\cite{ZS02PRE,nonparacrash}. It consists in a
generalization of the $q$-analysis \cite{E97PLA,EE97PRL}. The
$(H,q)$-derivative is defined by the following formula
\begin{equation}
 D_q^H f(x) \stackrel{\triangle}{=}\frac{f(x)-f(qx)}{(x-qx)^H}~,
 \label{Eq:HqD}
\end{equation}
such that $D_q^{H=1} f(x)$ recovers the standard $q$-derivative
$D_q f(x)$. For a power law function $f(x) = B x^m$, $D_q^{H=m} [B
x^m] = B (1-q^m)/(1-q)^m$ is constant. In a nutshell, the
$(H,q)$-analysis performs a kind of fractal derivative which is
particularly sensitive to the presence of log-periodicity. The
index $q$ refers to the discrete scaling ratio used in the
definition of the fractal $q$-derivative and $H$ is an exponent
used to rescale the $q$-derivative. Scanning $q$ and $H$ provides
an important test of the robustness of the log-periodic structure.
A Lomb periodogram analysis \cite{Numrec} of the
$(H,q)$-derivative allows one to detect the presence of
log-periodicity and assess its significance level.

\begin{figure}
\begin{center}
\epsfig{file=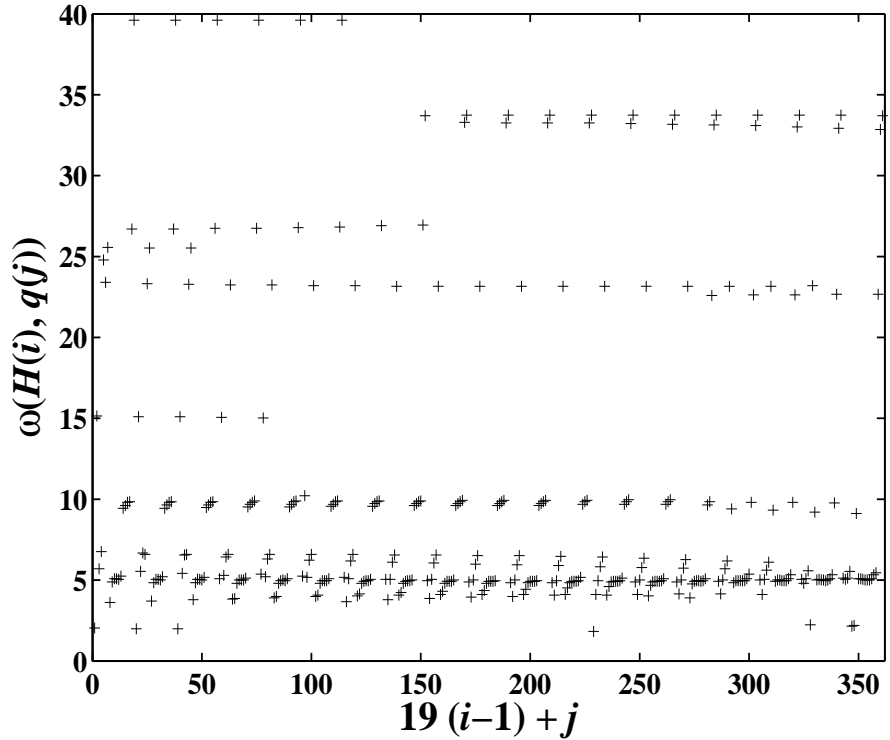,width=12cm, height=10cm}
\end{center}
\caption{Angular log-frequency $\omega$ of the most significant
Lomb peak in each lomb periodogram in the $(H,q)$-analysis of
$I(t)$. The analysis was performed on a $19\times 19$ grid with
$H$ ranging from -0.9 to 0.9 evenly and $q$ varying from 0.05 to
0.95 evenly. Each node $(i,j)$ in the grid corresponds to a pair
$(H(i),q(j))$, which is converted to a number in this plot through
a one-to-one map: $(i,j)\to 19\times(i-1)+j$. Each cross
corresponds to a pair of $(H,q)$. Note that $W(H(7),q(19))=50.4$
is not plotted in the figure for the sake of better presentation.}
\label{Fig:HqAW}
\end{figure}

The analysis was performed on a $19\times 19$ grid with $H$
ranging from $-0.9$ to $0.9$ evenly and $q$ varying from $0.05$ to $0.95$
evenly. Each node $(i,j)$ in the grid corresponds to a pair
$(H(i),q(j))$, which is converted to a single number
through the one-to-one map: $(i,j)\to 19\times(i-1)+j$. Figure
\ref{Fig:HqAW} uses this number as the abscissa
to show the corresponding angular log-frequency $\omega$ of the
most significant Lomb peak in each Lomb periodogram for each of
the scanned $(H,q)$ pairs of the $(H,q)$-derivative of the net foreign
capital inflow $I(t)$ into the USA. Each cross corresponds to a pair of
$(H,q)$.

The majority of the $\omega$'s cluster
around $\omega_1=5$, which is consistent with
the fundamental log-frequency extracted from the fit with equation
(\ref{Eq:It}). Furthermore, one can observe clusters of $\omega$'s
at integer multiples $\omega_2=10$ and $\omega_3=15$
of this fundamental angular log-frequency. Some higher-order
multiples $\omega_5 \approx 25$, $\omega_7 \approx 35$ and
$\omega_{10} \approx 50$ (not shown) are also clearly visible.
The deviations from the exact integer multiples are mostly due to
small $q$'s which are most sensitive to finite-size effects.

The $(H,q)$-analysis confirms the presence of a
very significant log-periodic structure in the bubble of the US
net capital inflow of foreign assets.

\clearpage

\clearpage

\clearpage

\clearpage

\clearpage

\clearpage

\clearpage

\end{document}